\documentclass[a4paper,fleqn,usenatbib]{mnras}

\usepackage{newtxtext,newtxmath}

\usepackage[T1]{fontenc}
\usepackage{ae,aecompl}

\usepackage{graphicx}	
\usepackage{amsmath}	
\usepackage{amssymb}	

\newcommand{\target}{PMN~J0948+0022}

\title[Radio jet structures of the $\gamma$-NLS1 PMN J0948+0022]{Radio jet structures at $\sim100$~parsec and larger scales of the $\gamma$-ray-emitting narrow-line Seyfert 1 galaxy PMN J0948+0022}

\author[Doi et al.]{
Akihiro Doi$^{1,2}$\thanks{E-mail: doi.akihiro@jaxa.jp (AD)}, 
Satomi Nakahara$^{1}$, 
Masanori Nakamura$^{3}$, 
Motoki Kino$^{4,5}$,
\newauthor
Nozomu Kawakatu$^{6}$, and  
Hiroshi Nagai$^{5}$
\\
$^{1}$The Institute of Space and Astronautical Science, Japan Aerospace Exploration Agency, 3-1-1 Yoshinodai, Chuou-ku, Sagamihara, Kanagawa 252-5210,\\ Japan\\
$^{2}$Department of Space and Astronautical Science, SOKENDAI (The Graduate University for Advanced Studies), 3-1-1 Yoshinodai, Chuou-ku, Sagamihara,\\ Kanagawa 252-5210, Japan\\
$^{3}$Academia Sinica Institute of Astronomy and Astrophysics, P.O. Box 23-141, Taipei 10617, Taiwan\\
$^{4}$Kogakuin University of Technology \& Engineering, Academic Support Center, 2665-1 Nakano, Hachioji, Tokyo 192-0015, Japan\\
$^{5}$National Astronomical Observatory of Japan, 2-21-1 Osawa, Mitaka, Tokyo 181-8588, Japan\\
$^{6}$National Institute of Technology, Kure College, 2-2-11, Agaminami, Kure, Hiroshima 737-8506, Japan\\
}

\begin{document}

\date{Accepted 2019 May 2. Received 2019 May 1; in original form 2018 November 17}

\pagerange{\pageref{firstpage}--\pageref{lastpage}} 
\pubyear{2019}

\maketitle

\label{firstpage}


\begin{abstract}
The narrow-line Seyfert~1 (NLS1) galaxy PMN~J0948+0022 is an archetype of gamma-ray-emitting NLS1s in active galactic nuclei~(AGNs). In this study, we investigate its radio structures using archival data obtained using the Karl G.~Jansky very large array~(VLA) and the very long baseline array~(VLBA).  
The new VLA images reveal an outermost radio emission separated by 9.1~arcsec.  Its resolved structure and steep spectrum suggest a terminal shock in a radio lobe energized by the jet from the PMN~J0948+0022 nucleus.  This large-scale radio component is found at almost the same position angle as that of the pc-scale jet, indicating a stable jet direction up to $\sim1$~Mpc.  Its apparent one-sidedness implies a moderate advancing speed ($\beta>0.27$).  The kinematic age of $1 \times 10^7$~year of the jet activity is consistent with the expected NLS1 phase of $\sim10^7$~year in the AGN lifetime.  
The VLBA image reveals the jet structure at distances ranging from $r=0.82$~milliarcsec to $3.5$~milliarcsec, corresponding to approximately 100~pc, where superluminal motions were found.  The jet width profile ($\propto r^{1.12}$) and flux-density profile ($\propto r^{-1.44}$) depending on the distance $r$ along the jet suggest that the jet kinetic energy is converted into internal energy in this region.  The jet is causally connected in a nearly conical streamline, which is supported by ambient pressure at 100-pc scales in the host galaxy of PMN~J0948+0022.        
\end{abstract} %


\section{INTRODUCTION}\label{section:introduction} 
Narrow-line Seyfert~1 (NLS1) galaxies are a subclass of active galactic nuclei~(AGNs) identified by their optical properties of a line ratio [\ion{O}{iii}]/H$\beta<3$~and Balmer lines that are only slightly broader than the forbidden lines (\citealt{Osterbrock:1985}) in the definition of full-width at half maximum FWHM(H$\beta)<2000$~km~s$^{-1}$ \citep{Goodrich:1989}.  
Detections in the $\gamma$-ray regime from a dozen NLS1s (\citealt{Abdo:2009,Foschini:2011,DAmmando:2012,Liao:2015,Yao:2015};\\ \citealt{Yang:2018,Paliya:2018,Lahteenmaki:2018,Paiano:2018}) have prompted great interest in their $\gamma$-ray production mechanisms and parent population.    
These $\gamma$-ray-emitting NLS1s ($\gamma$-NLS1s) exhibit blazar-like properties, double-peaked broad-band spectral energy distributions~(SEDs; e.g. \citealt{Abdo:2009}), highly variable flat/inverted radio spectra \citep[e.g.][]{Angelakis:2015,Lahteenmaki:2017}, extremely high brightness temperatures \citep[e.g.][]{Zhou:2003,Doi:2006}, one-sided core--jet morphology \citep[e.g.][]{Doi:2006,Giroletti:2011}, superluminal motions \citep[e.g.][]{DAmmando:2013,Fuhrmann:2016} and variable optical polarizations \citep[e.g.][]{Itoh:2013}, which are believed to be a result of beaming/light-travel-time effects due to a relativistic jet aligned close to our line of sight \citep{Blandford:1979}.  
NLS1s showing very large radio-loudness $R>100$\footnote{The radio loudness $R$ is defined as the ratio of 5~GHz radio to {\it B}-band flux densities, with a threshold of $R=10$ separating the radio-loud and radio-quiet objects \citep{Kellermann:1989}} also exhibit blazar-like properties \citep[e.g.][]{Yuan:2008}.  The jet powers of blazar-like NLS1s are similar to those of BL~Lac objects, even those with prominent optical emission lines, or the low-power end of flat-spectrum radio quasars \citep[FSRQs; e.g.][]{Foschini:2015}.  
On the other hand`, it has been suggested that a fraction of radio-loud NLS1s ($R>10$) shows radio properties similar to compact steep spectrum radio sources \citep{Komossa:2006,Doi:2011,Doi:2016a,Gu:2015}, indicating sufficiently large jet powers in the non-beamed population.  
Most NLS1s ($\sim93$\%) are radio-quiet \citep[$R<10$;][]{Zhou:2006}.  Detections with brightness temperatures of $>10^7$~K in several radio-quiet NLS1s also indicate a nonthermal jet origin \citep{Middelberg:2004,Giroletti:2009,Doi:2013a,Doi:2015}.

\target~($z=0.585$) was discovered in the Sloan Digital Sky Survey~(SDSS) early data release \citep{Williams:2002,Zhou:2003} as a very radio-loud NLS1, showing blazar-like properties such as highly variable radio emission \citep{Zhou:2003}, a very compact flat/inverted-spectrum radio core, and a one-sided morphology revealed by milliarcsecond~(mas) resolutions using very-long-baseline interferometry \citep[VLBI;][]{Doi:2006}.  The viewing angle $\theta$ and Doppler factor $\delta$ of the jet were mildly restricted by the radio observations.  The extremely high brightness temperatures implied $\delta \ga 2.5$ \citep{Zhou:2003} or $\delta > 5.5$ according to the rapid variability and $\delta > 2.7$ and $\theta<22\degr$ according to the compactness of the radio core \citep{Doi:2006}. 
  
The first detection of GeV $\gamma$-ray emission from NLS1s was made in \target\ by the large-area telescope on board the {\it Fermi Gamma-ray Space Telescope} satellite \citep[{\it Fermi} LAT;][]{Abdo:2009a}.   
After the $\gamma$-ray detection, \target~ has been extensively explored by several multiwavelength monitoring campaigns \citep{Abdo:2009b,Foschini:2011a,Foschini:2012,DAmmando:2014,DAmmando:2015}.  
The SEDs can be reproduced via the radiative processes of synchrotron, synchrotron self-Compton and inverse-Compton scattering of external infrared radiation from a dust torus by assuming a relativistic jet with a bulk Lorentz factor of $\Gamma=10$ and a viewing angle of $\theta=6\degr$ \citep{Abdo:2009a}, $\Gamma=11$--$16$ and $\theta=3\degr$ \citep{Foschini:2011a,Foschini:2012} or $\Gamma=16$--$30$ \citep{DAmmando:2015}. 
PMN~J0948+0022 exhibits behaviour similar to those of other FSRQs in terms of its SED, even though its jet power is at the lower limit for FSRQs \citep{Abdo:2009a,DAmmando:2015}.  
Its variable radio spectrum has been intensively monitored \citep{Angelakis:2015,Lahteenmaki:2017}, and a variability Doppler factor of $\delta=8.4$ has been obtained \citep{Angelakis:2015}.  
Violent intranight optical variability and variable optical polarization similar to the characteristics observed in the BL~Lac AGN class have been observed; the origin of this behaviour could be a nearly pole-on viewed relativistic jet \citep{Liu:2010,Itoh:2013,Maune:2013,Paliya:2013a,Eggen:2013}.  
Its ultraviolet bump is ascribed to thermal emission from an accretion disk as in FSRQs \citep{Abdo:2009b,Foschini:2011a,DAmmando:2015}.  
The origin of the soft X-ray excess observed in its intermediate states is still under debate; \citet{DAmmando:2014} and \citet{Bhattacharyya:2014} have advocated an accretion-disk/corona system origin as is usual for many radio-quiet NLS1s \citep[e.g.][]{Leighly:1999a,Grupe:2010}, while the power-law component above 2 keV most likely originates in the jet.  

Therefore, \target~is considered an archetype of $\gamma$-NLS1s and has been extensively investigated.  However, its jet structures have not been well-determined via direct radio imaging.  This may be because \target~is a relatively distant object.  
\target~exhibits only a compact core-jet morphology, elongated at position angles between $20\degr$ and $40\degr$, for observations between frequencies of 5 and 22~GHz in VLBI images \citep{Doi:2006,Giroletti:2011}.  Other VLBI images have been obtained at 15.3~GHz in the Monitoring of Jets in AGNs with very long baseline array~(VLBA) experiments \citep[MOJAVE;][]{Lister:2005}, and have been reported via a series of multiwavelength monitoring campaigns \citep[e.g.][]{Foschini:2011a}.  The brightness temperature of its compact core exceeds $6\times10^{12}$~K, with an average fractional polarisation of 0.8\% \citep{Foschini:2012}.  
A low-brightness jet in \target\ extends up to only $\sim3$~mas \citep{Pushkarev:2017} with an apparent opening angle of $21\degr$ \citep{Foschini:2012}.   
Two superluminal motions of $(11.5 \pm 1.5)c$ and $(5.5 \pm 1.9)c$ have been discovered for \target\ \citep{Lister:2016}.  The jet-width profile dependence of the radial distance from the core has been measured as a power-law index of 1.83 in a stacked MOJAVE image using data from 17~epochs \citep{Pushkarev:2017}.  The further extension of the jet structure is unknown at mas resolutions because of being unresolved at 1.7 and 2.3~GHz \citep{Doi:2006}.  
At arcsecond resolutions, only one very large array~(VLA) image from the Faint Images of the Radio Sky at Twenty-Centimeters~\citep[FIRST;][]{Becker:1995} using the B-array configuration at 1.4~GHz has been presented \citep{Doi:2012}; a possible two-sided elongation in the north--south direction, which was barely distinguishable at FIRST's $\sim6\arcsec$ resolution, suggests an arm length of $\sim50$~kpc in the projected distance.  A quasi-simultaneous radio spectrum between the frequencies of 1.4 and 15~GHz was obtained using the VLA B-array configuration; however, images were not presented due to the unresolved morphology \citep{Doi:2006}.  
The postage stamp server\footnote{\url{https://www.cv.nrao.edu/nvss/postage.shtml}} for the National Radio Astronomy Observatory~(NRAO) VLA Sky Survey~\citep[NVSS;][]{Condon:1998} returns radio images at 1.4~GHz using the D-array configuration.  A second radio source in the northeast direction is associated with PMN~J0948+0022 \citep{Komossa:2006}; its physical origin is unknown.  
Accordingly, traces of the energy transfer from the central engine to the outermost environment have not been thoroughly investigated for the $\gamma$-NLS1 \target.

In the present study, we investigate the radio jet structures at approximately 100~pc and larger scales of \target~on the basis of the radio images obtained using VLBA and the Karl G.~Jansky VLA, respectively.  These data contribute to a more comprehensive understanding of the jet properties downstream of the gamma-ray dissipation region.  
The rest of this paper is organized as follows.  In Section~\ref{section:data}, we describe the data and imaging procedures.  The results are presented in Section~\ref{section:results}.  We discuss the jet structures of this $\gamma$-NLS1 in Section~\ref{section:discussion}.  Finally, we summarize the study in Section~\ref{section:summary}.  
Throughout the paper, we assume a $\Lambda$CDM cosmology with $H_0=70.5$~km~s$^{-1}$~Mpc$^{-1}$, $\Omega_\mathrm{M}=0.27$ and $\Omega_\mathrm{\Lambda}=0.73$.  At the distance of \target, an angular size of 1\arcsec corresponds to 6.6~kpc in projection.

\begin{table*}
\begin{minipage}{170mm}
\begin{center}
\caption{Data, image parameters and source properties of component N.}
\label{table:data}
\begin{tabular}{lllccrcccc}
\hline\hline
Project	&	Date	&	Array	&	$\nu$	&	$\theta_\mathrm{maj}^\mathrm{beam} \times \theta_\mathrm{min}^\mathrm{beam}$	&	PA$^\mathrm{beam}$	&	$\sigma$	&	$F_\nu^{\rm N}$			&	$\theta_\mathrm{maj}^\mathrm{N} \times \theta_\mathrm{min}^\mathrm{N}$			&	PA$^\mathrm{N}$			\\
	&		&		&	(GHz)	&	(\arcsec $\times$ \arcsec)	&	(\degr)	&	(mJy beam$^{-1}$)	&	(mJy)			&	(\arcsec $\times$ \arcsec)			&	(\degr)			\\
(1)	&	(2)	&	(3)	&	(4)	&	(5)	&	(6)	&	(7)	&	(8)			&	(9)			&	(10)			\\
\hline
NVSS	&	27 Feb.~1995	&	VLA-D	&	1.40	&	$45 \times 45$	&	\ldots	&	0.450	&	\ldots			&	\ldots			&	\ldots			\\
FIRST	&	10~Aug.--19~Sep.~1998 	&	VLA-B	&	1.40	&	$6.4 \times 5.4$	&	$0$	&	0.145	&	\ldots			&	\ldots			&	\ldots			\\
15A-337	&	14~Aug.~2015	&	VLA-A	&	1.64	&	$1.05 \times 0.99$	&	$-54$	&	0.101	& $	3.36	\pm	0.15	$ & $	0.69	\times	0.38	$ & $	66	\pm	14	$ \\
AK360	&	07~May~1994	&	VLA-AB	&	4.86	&	$1.02 \times 0.75$	&	$84$	&	0.244	& $	1.03	\pm	0.37	$ & $	1.33	\times	0.74	$ & $	76	\pm	20	$ \\
AG688	&	24~Dec.~2004	&	VLA-A	&	8.46	&	$0.31 \times 0.25$	&	$21$	&	0.057	& $	0.88	\pm	0.10	$ & $	0.46	\times	0.22	$ & $	91	\pm	14	$ \\
TSKY0001	&	25~Feb.~2018	&	VLA-A	&	10.00	&	$0.30 \times 0.28$	&	$-12$	&	0.039	& $	0.61	\pm	0.09	$ & $	0.52	\times	0.23	$ & $	73	\pm	8	$ \\
15B-210	&	02--30~Nov.~2015$\dagger$	&	VLA-D	&	33.19 	& 	$1.93 \times 1.83$	&	$-2$	&	0.011	& $	0.18	\pm	0.01	$ & $	<0.1			$ & 	\ldots			\\

\hline
\end{tabular}
\end{center}

\begin{footnotesize}
Note.~The columns are as follows: (1) project ID; (2) observation date; (3) array configuration; (4) observing frequency; (5) beam size; (6) position angle of the beam's major axis; (7) image noise; (8) flux density of component N; (9) deconvolved size and (10) position angle of the deconvolved source structure.  \\
$\dagger$ Data from eight observations ($\mathrm{MJD} =$ 57328, 57334, 57339, 57340, 57345, 57346, 57349 and 57356) were concatenated.   
\end{footnotesize}
\end{minipage}
\end{table*}

\section{Data and Imaging}\label{section:data}
We retrieved VLA archival data, which included observations of \target~from the National Radio Astronomy Observatory~(NRAO) archives.  Data that were expected to provide high-sensitivity images ($\la0.1$~mJy~beam$^{-1}$) at moderately high angular resolutions ($\la 1\arcsec$) were selected.  Our aim was to resolve the low-brightness radio structure that had been seen as only a slightly elongated structure in the FIRST image at a resolution of $\sim6\arcsec$ \citep{Doi:2012}.

\subsection{VLA data}\label{section:JVLAimages}
The data with project codes of 15A-337, TSKY0001, and 15B-210 were obtained using VLA and the Wideband Interferometric Digital ARchitecture~(WIDAR) correlator~(Table~\ref{table:data}).  
The 15A-337 observation was carried out using the A-array configuration and dual polarization at a centre frequency of $\sim1.5$~GHz with a bandwidth of 1~GHz (1007.5--2031.5~MHz), consisting of 16~spectral windows~(spw) with 64~MHz bandwidth each.  The total on-source time on \target\ was approximately 700~s.  
The TSKY0001 observation was carried out as part of the VLA Sky Survey~(VLASS\footnote{\url{https://science.nrao.edu/science/surveys/vlass}}) using the AnB-array configuration and dual polarization at a centre frequency of $\sim10$~GHz with a bandwidth of 5~GHz (7503--12503~MHz), including 49~spw with 128~MHz bandwidth each.  The on-source time on \target\ was 55~s.  
The 15B-210 observation was carried out using the D-array configuration and dual polarization at a centre frequency of $\sim33$~GHz with a bandwidth of 2~GHz (32188--34172~MHz), including 16~spw with 128~MHz bandwidth each.  The on-source time was approximately 15,000~s on \target.  
We reduced and analysed these VLA data using the Common Astronomy Software Applications \citep[CASA;][]{McMullin:2007} version 4.7 by following the standard expanded VLA data reduction guidelines for continuum observations.    
Radio frequency interference (\footnote{We also attempted to use the auto-flagging algorithm ``rflag'' within the {\tt CASA} task {\tt flagdata} instead of manual channel flagging.  However, the results were not significantly improved.}) with amplitudes significantly larger than the correlated fluxes was observed in the L-band and X-band.  We conducted data flagging on approximately $\sim74$\% and $\sim29$\% of the frequency channels in the 15A-337 and TSKY0001 data, respectively.  
We created final images in Stokes~I using the {\tt CASA} task {\tt tclean} with multifrequency synthesis and natural weighting (on 15A-337 and 15B-210) or Briggs weighting \citep[${\rm robust}=0$;][on TSKY0001]{Briggs:1995}.  In the 15A-337 data, several bright outlier sources in the primary beam that needed to be modelled were simultaneously imaged to minimize the sidelobes in the area of interest.  In addition, we developed an image from the uv-tapered visibilities with a resolution of $6\arcsec$, which is equivalent to that of the FIRST image.

%
From the NRAO archives, we retrieved two archival datasets of historical VLA observations of \target.
AK0360 and AG688 were obtained with the AB-array configuration at $4.9$~GHz for 50~s on-source and with the A-array configuration at $8.5$~GHz for approximately 600 s on-source on \target~(Table~\ref{table:data}).    
These observations were performed in full-polarization mode with two IF channels of 50~MHz (100~MHz in total).  
Data reduction was performed using the Astronomical Image Processing System~({\tt AIPS}; \citealt{Greisen:2003}) according to the standard procedures for VLA continuum observations.  Final calibrations were iteratively performed by CLEAN deconvolution and self-calibration using {\tt difmap} software \citep{Shepherd:1994}.  We obtained the final images in Stokes~I with natural weighting.  

We also retrieved a total intensity contour map of NVSS \citep{Condon:1998} obtained at 1.4~GHz using the D-array configuration (at an angular resolution of $45\arcsec$), to investigate the radio environment around \target~in a wide field of view.  The image noise was $\sigma = 0.45$~mJy~beam$^{-1}$.

\subsection{VLBA data}\label{section:VLBAimages}
We retrieved an archival image of \target~observed at 15.35~GHz using the VLBA from the MOJAVE project \citep{Lister:2005} online page\footnote{\url{https://www.physics.purdue.edu/MOJAVE/sourcepages/0946+006.shtml}}.  
The image is a total intensity map generated by stacking the data acquired from 28~May 2009 to 30~July 2013 (17 epochs in total) and convolved using a restored circular beam of $0.82$~mas \citep{Pushkarev:2017}.

\begin{figure*}
\begin{minipage}{170mm}
\includegraphics[width=\linewidth]{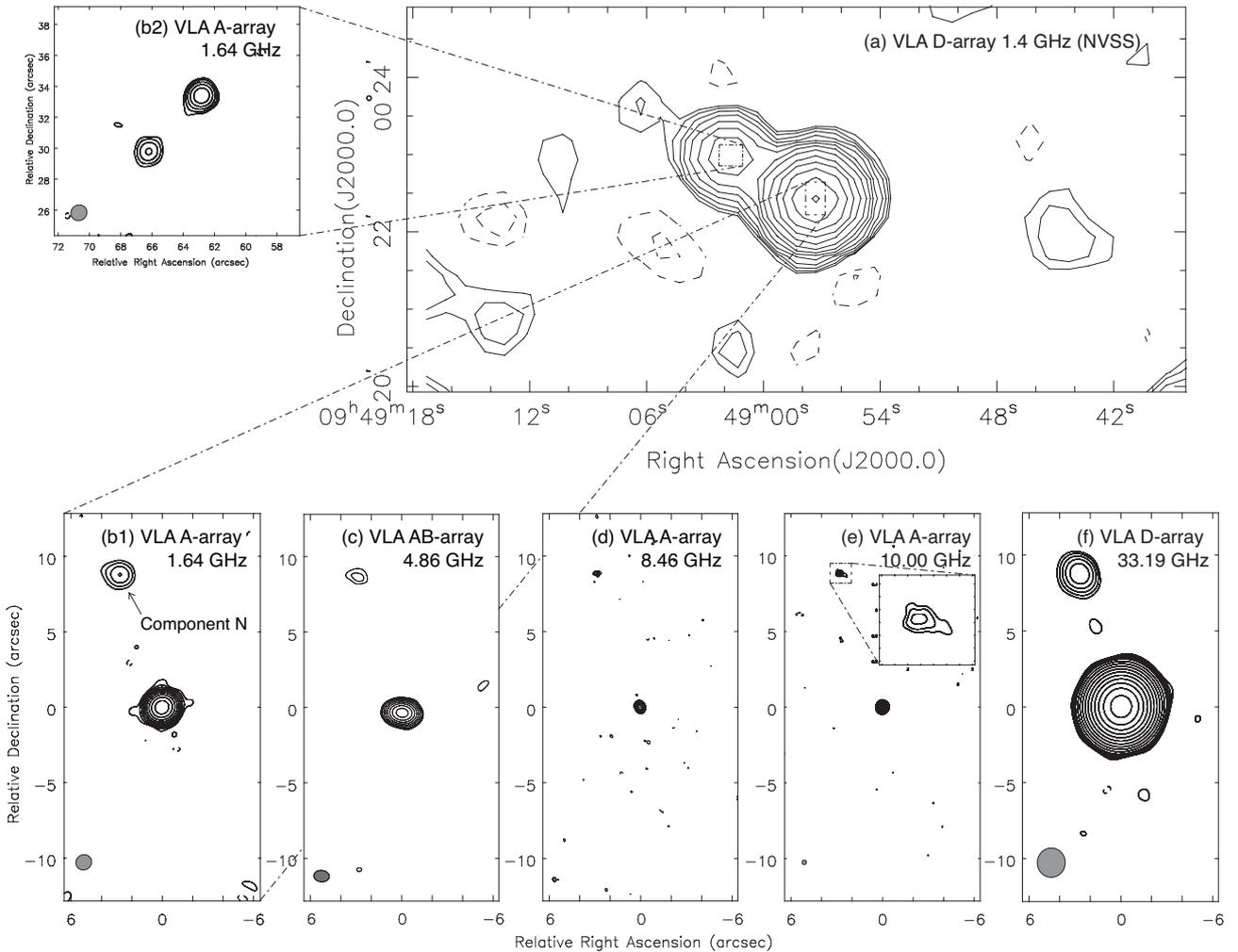}
\caption{Radio images of PMN~J0948+0022 at arcsecond scales:
(a)~the wide-field image of NVSS at 1.4~GHz centred at the position of PMN~J0948+0022; 
(b1) the VLA A-array image in the central region at 1.64~GHz; (b2) the VLA A-array image focused on the separated region centred at ($\rm{\Delta RA, \Delta Dec})\sim(65\arcsec,32\arcsec)$ from PMN~J0948+0022; 
(c) the VLA AB-array image of the central region at 4.86~GHz;
(d) the VLA A-array image of the central region at 8.46~GHz;
(e) the VLA A-array image of the central region at 10.00~GHz, the sub-image is the magnification of component `N' and
(f) the VLA D-array image of the central region at 33.19~GHz.  
The contour levels are separated by a factor of 2, beginning at $3\sigma$ of the rms image noise.  A $3\sigma \times \sqrt{2}$ contour is added in panels (c), (d) and (e); $3\sigma \times \sqrt{2}, 2\sqrt{2}$ contours are added in panel (f).  The beam size is illustrated as a filled ellipse in the lower left of each panel for the central region.  
}
\label{figure:VLAimages}
\end{minipage}
\end{figure*}

\begin{figure}
\begin{center}
\includegraphics[width=\linewidth]{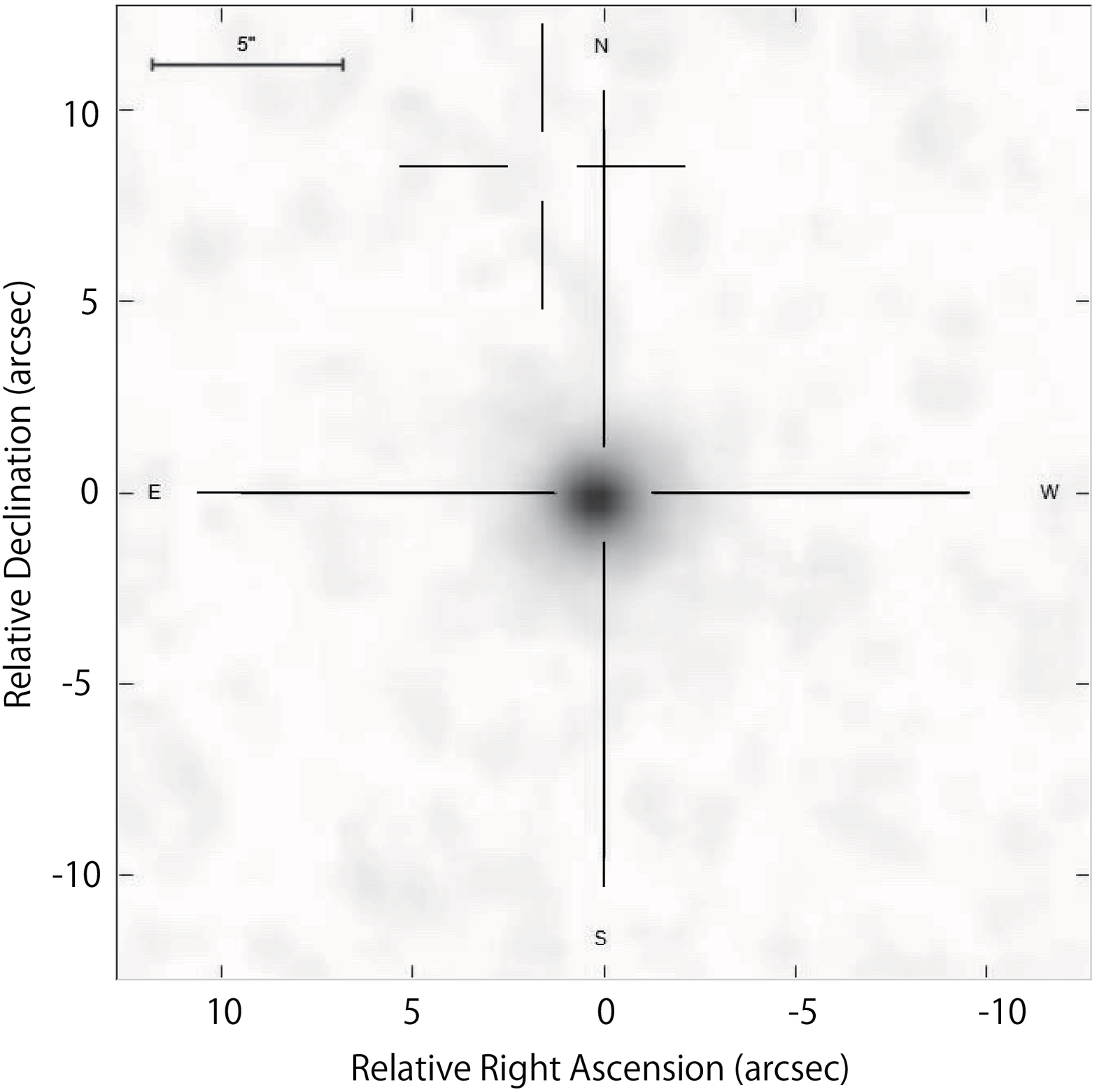}
\end{center}
\caption{SDSS DR9 image of PMN~J0948+0022.  The cross at the image centre indicates the position of PMN~J0948+0022.  The other cross in the north--northeast position indicates the location of the radio component N.}   
\label{figure:SDSSimage}
\end{figure}

\section{RESULTS}\label{section:results}

\subsection{Radio emissions at large scales}\label{section:results:kpcscaleradioemission}
 
Figure~\ref{figure:VLAimages} shows the VLA images in addition to the NVSS image for \target\ at arcsecond resolutions.  The image parameters are shown in Table~\ref{table:data}.  
 
In the NVSS image (Figure~\ref{figure:VLAimages}(a)), at first glance, it appears that the radio jet reaches the northeast.  
This second radio emission, at a separation of $71\arcsec$ from \target~nucleus, is catalogued as NVSS J094901+002258 \citep{Condon:1998} and FIRST~J094901.5+002258 \citep{Helfand:2015}.  No optical counterpart has been found in that sky position on the basis of the NASA/IPAC Extragalactic Database~(NED)\footnote{\url{http://ned.ipac.caltech.edu/}}, as previously mentioned by \citet{Komossa:2006}.   
In the VLA image at a higher angular resolution and 1.64~GHz (Figure~\ref{figure:VLAimages}(b2)), the emission clearly shows a double-source morphology.

The five VLA datasets provide radio images in the inner region, as shown in Figures~\ref{figure:VLAimages}(b1), \ref{figure:VLAimages}(c), \ref{figure:VLAimages}(d), \ref{figure:VLAimages}(e) and \ref{figure:VLAimages}(f).  
We detected an emitting component, hereafter called the component `N', in the north region of \target~nucleus in all VLA images at high angular resolutions.      
The component N is separated by $9.1\arcsec$, corresponding to 60~kpc in the projected distance, in the direction of $PA = 18\fdg2$ from \target~nucleus.       
We retrieved an optical image from the 9th Data Release of the Sloan Digital Sky Survey~(SDSS) from the SDSS Sky server\footnote{\url{http://skyserver.sdss.org/dr9/}} (Figure~\ref{figure:SDSSimage}).   No optical counterpart was found at the position of the radio component N.   The morphology of the host galaxy of \target~is unclear.

\begin{figure}
\includegraphics[width=\linewidth]{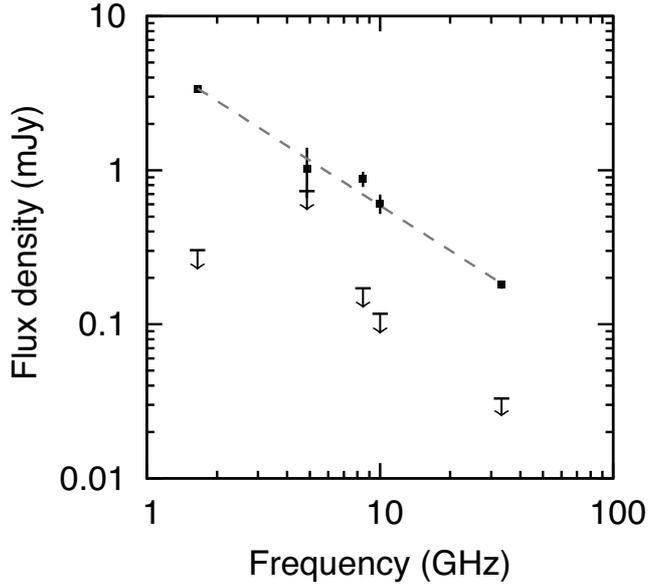}
\caption{Radio spectra of the large-scale radio emissions.  The filled symbols represent the measured flux densities for component N.  The dashed line represents a fitted power-law spectrum with an index of $-0.97 \pm 0.03$.  The downward arrows represent the upper limits ($3\sigma$) for the southern emission.}     
\label{figure:kpcspectrum}
\end{figure}

Image analyses using the {\tt CASA} task {\tt IMFIT} indicated that the component N is significantly resolved but moderately compact, compared to the beam sizes.  The size measurements showed similar results at all frequencies, except for 33.19~GHz (Table~\ref{table:data}), where the component was unresolved ($<0\farcs1$).  In the VLA images at 1.64 and 10.00~GHz, the source is elongated at $PA \sim 70\degr$.  The same structure of the component N was also clearly found in the VLA images at 4.86 and 8.46~GHz.  The apparent opening angle was calculated to be $\sim9\degr$ based on its deconvolved size ($\sim0\farcs7$ in the 1.64-GHz VLA image) and its separation from the central component, which was definitely unresolved ($<0\farcs04$).     
The flux densities of the component N were measured and exhibited a steep spectrum with a power-law index of $\alpha \sim -1.0$ (Figure~\ref{figure:kpcspectrum}) in the definition of $F_\nu \propto \nu^\alpha$, where $F_\nu$ is the flux density at the frequency $\nu$.

Conversely, no radio emission signature was seen southward in any of the new VLA images (Figures~\ref{figure:VLAimages}.  
However, the VLA FIRST image at a lower angular resolution ($6\farcs4 \times 5\farcs4$, B-array configuration at 1.4~GHz) and a worse sensitivity ($\sim0.15$~mJy~beam$^{-1}$) indicated a two-sided elongation in the north--south direction \citep{Doi:2012}.  If this can be attributed to the resolution effect, we would have to assume considerable asymmetry in the compactness for the two emitting regions.   However, our 1.64-GHz VLA image with an equivalent resolution by uv-taper (Section~\ref{section:JVLAimages}) did not show any signal in the south region ($1\sigma=0.31$~mJy~beam$^{-1}$).  
For confirmation, we re-examined the FIRST image.  
Image-based model fitting using the {\tt AIPS} task {\tt JMFIT} with three model components resulted in flux densities of 
$2.23 \pm 0.25$~mJy and 
$0.81 \pm 0.22$~mJy 
for the northern and southern emissions, respectively.  The signal-to-noise ratio ${\rm SNR} = 3.6$ suggests a marginal detection for the southern emission.  
We also analysed the original visibility data of the FIRST mosaicing observations around \target.  Finally, an image similar to the FIRST image was obtained; the northern emission is clearly seen and the southern one once again looks marginal.  We can recognize sidelobe patterns at position angles of $\pm20\degr$ and $\pm 160\degr$, which potentially affect the regions close to bright emissions.  Such characteristic sidelobes may originate in the snapshot-style observations obtained using the {\sf Y}-shaped antenna configuration of VLA.  

We conclude that the component N was definitely detected; however, we have not yet confirmed the southern counterpart at the VLA sensitivities with the WIDAR correlator.  The deepest upper limit is given by the VLA data at 1.64~GHz if a spectral index of $-1$ derived from the component N is assumed.     
Because N was resolved into a convolved size equivalent to 1.25 times the beam area (Table~\ref{table:data}), the upper limit of the flux density for a putative counter component is defined as $1.25 \times 3\sigma$, where $\sigma$ is the image noise ($\sigma = 0.10$~mJy~beam$^{-1}$); as a result, the flux ratio of the northern to southern emissions is $R_{\rm F} > 8.9$.

\subsection{The pc-scale jet structure}\label{section:results:pcscalejetstrcture}
\begin{figure}
\includegraphics[width=\linewidth]{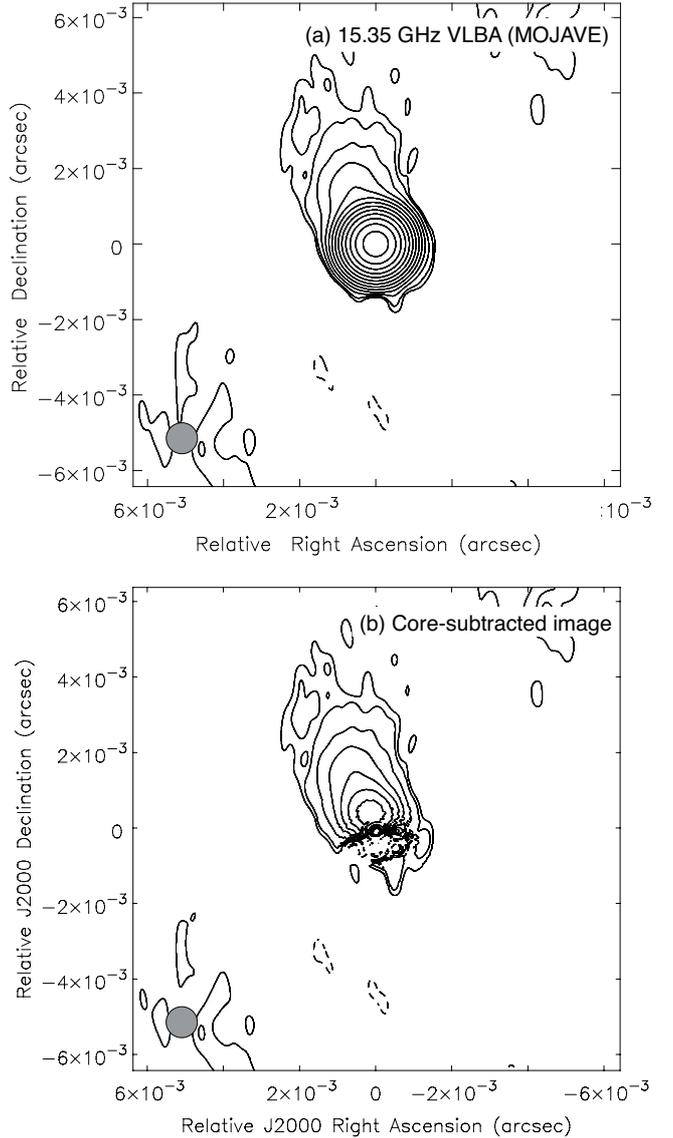}
\caption{(a)~MOJAVE VLBA image of PMN~J0948+0022 at pc scales at 15.35~GHz.  (b)~The core-subtracted image.  The contour levels are separated by a factor of 2, beginning at $3\sigma$ of the rms image noise ($1\sigma = 50\ \mu \rm{Jy}$); a $2-\sigma$ contour is added to trace the more diffuse emission.  The image is convolved using a restored circular beam of $0.82$~mas.}   
\label{figure:VLBAimage}
\end{figure}

Figure~\ref{figure:VLBAimage}(a) is a stacked VLBA image from MOJAVE, showing the pc-scale jet structure of \target.  
The morphology is highly core-dominated, and the jet intensity close to the core is severely contaminated by the Gaussian tail of the convolution for an unresolved strong emission at the nucleus.     
We developed an only-jet image via an image-based subtraction of a point source model with a flux density that equalled the peak intensity in the original image (Figure~\ref{figure:VLBAimage}(b)). 

\begin{figure}
\includegraphics[width=\linewidth]{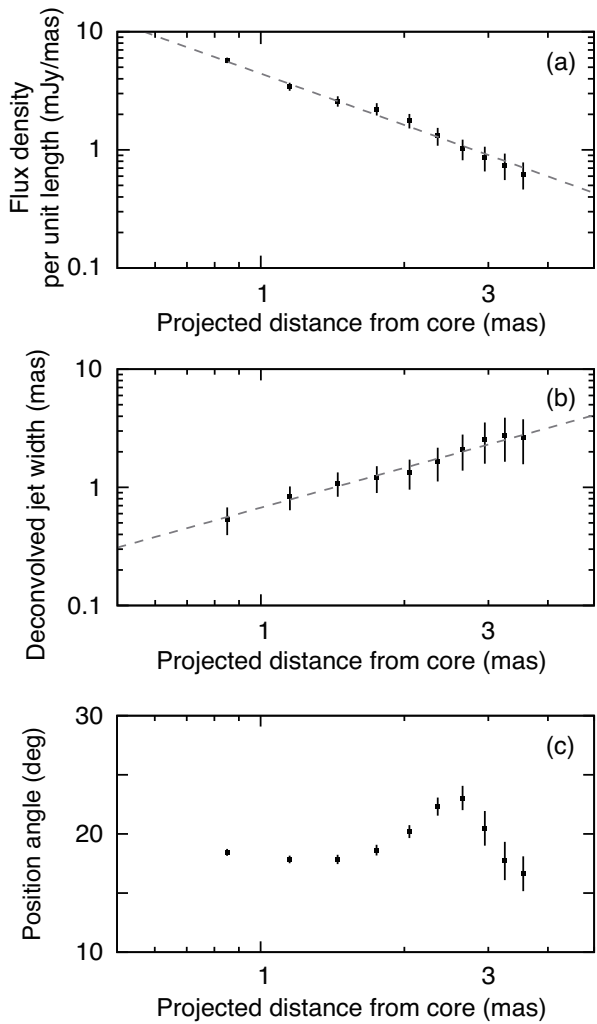}
\caption{Pc-scale jet structures of PMN~J0948+0022 based on the core-subtracted image.  (a) The profile of the flux density per unit length with distance from the core.  The dashed line shows the fitted profile: $b=-1.44 \pm 0.01$, where $F_r^\prime \propto r^b$.  (b)~The profile of the deconvolved jet width (diameter).  The dashed line represents the fitted jet width profile: $k=1.12 \pm 0.03$, where $w_r \propto r^k$.  (c)~The profile of the position angle of the ridge line of the jet.}   
\label{figure:jetprofile}
\end{figure}

To investigate the jet structure, we conducted a pixel-based image analysis following the method used by researchers such as \citet{Nakahara:2018}, in which the transverse profile of the intensity distribution of the jet at different distances was sliced and modelled using a Gaussian profile.  This method assumes a Gaussian profile for the intrinsic jet intensity distribution in the transverse direction.  This assumption is reasonable for the jet of \target\ because it does not exhibit limb-brightening like those of 3C~84 and M87~\citep[e.g.][]{Nagai:2014,Asada:2016}.    
Using the {\tt AIPS} task {\tt SLICE}, we sliced the jet transversely at different distances (every 0.15~mas, corresponding to $\sim1/5$ of the synthesized beam, e.g. \citealt{Lobanov:2005}) from the core along the putative position angle of the jet $PA=20\degr$.    
Assuming a convolution beam of $\theta_{\rm beam}=0.8195$~mas (described in the FITS header), we determined the deconvolved structure of the jet.  Consequently, we obtained the profiles of the jet intensity ($I_r$; not shown), the deconvolved FWHM of the jet width ($w_r$; Figure~\ref{figure:jetprofile}(b)) and the ridge line (Figure~\ref{figure:jetprofile}(c)).  

The determined jet intensity profile continues up to 3.5~mas from the core above the $3\sigma$ image noise.  We calculated the flux density per unit length (1~mas) along $PA=20\degr$ to be $F_r^\prime = I_r (w_r^2 + \theta_{\rm beam}^2)^{0.5}/\theta_{\rm beam}^2$ \citep{Nakahara:2018}, as shown in Figure~\ref{figure:jetprofile}(a).  
The flux density per unit length was fitted to a single power-law function, resulting in the index $b=-1.44 \pm 0.02$, where $F_r^\prime \propto r^b$.  

The deconvolved jet width profile was fitted, resulting in the index $k=1.12 \pm 0.03$ and scale factor $w_0= 0.67 \pm 0.01$~mas at $r=1$~mas, where $w_r = w_0 r^k$.  The apparent (full-)opening angle of the jet was derived to be $41\degr \pm 1\degr$ on average.   \citet{Pushkarev:2017} reported a power-law index of $k=1.83 \pm 0.08$ for \target~via a similar analysis on the identical image data but using the original stacked (i.e. not core-subtracted) image; we also obtained nearly the same value ($1.84 \pm 0.07$) by using the original stacked image.  This discrepancy indicates that the latter result was affected by contamination from a strong core emission, which certainly includes the jet-width profile in unresolved regions and leaks into the downstream through the Gaussian tail of the convolution function.  In the present paper, we adopt $k=1.12 \pm 0.03$ based on the core-subtracted image.  

The determined jet ridge structure appears from a corresponding PA of $\sim23\degr$ approaching $\sim17\deg$.  The average is $PA = 18\fdg6 \pm 0\fdg4$, which is consistent with the direction of the northward large-scale radio component N detected in the VLA images ($PA = 18\fdg2$; Section~\ref{section:results:kpcscaleradioemission}).

\section{DISCUSSION}\label{section:discussion}
\subsection{Jet activity in large scales}\label{section:discussion:maxmum_extent}
The apparent one-sided radio morphology seen in the NVSS image (Figure~\ref{figure:VLAimages}(a)) can be examined using the new VLA image at 1.64~GHz at a higher angular resolution.    
We found the emitting region that is separated by $71\arcsec$, corresponds to 470~kpc in the projected distance from \target.   \citet{Komossa:2006} previously mentioned that no optical counterpart could be identified at this radio position. 
The morphology is revealed to be a double source (Figure~\ref{figure:VLAimages}(b2)), strongly indicating another radio galaxy.     
Therefore, the apparent one-sided radio morphology in the NVSS image is caused by a neighbouring background radio source.

In the inner region, we detected the northern component N in all VLA images at high angular resolutions.  
The component N showed the following properties: (1)~no optical counterpart, (2)~a separation of $9.1\arcsec$ (corresponding to 60~kpc in the projected distance) in the direction ${\rm PA} \sim 18\degr$ from \target~nucleus, (3)~compact but significantly resolved in high-angular-resolution images, (4)~elongated at ${\rm PA} \sim 70\degr$, a very different direction with respect to that of the position angle, and (5)~a steep radio spectrum ($\alpha = -1.0$), as presented in Section~\ref{section:results:kpcscaleradioemission}.  These radio properties suggest a terminal shock in a radio lobe emanating from \target~nucleus in the jet direction (${\rm PA} \sim 18\degr$).  The radio morphology suggests an edge-brightened radio structure reminiscent of Fanaroff--Riley class-II radio galaxies \citep{Fanaroff:1974}, though no signature of a bridging structure between N and the nucleus was found at our limited image sensitivities.  

FR~II-like morphology has been more clearly discovered in several radio-loud NLS1s (PKS~0558-504, \citealt{Gliozzi:2010}; SDSS~J120014.08$-$004638.7, \citealt{Doi:2012}; J0953+2836, J1435+3131, J1722+5654, \citealt{Richards:2015}; J0814+5609, \citealt{Berton:2018} and SDSS~J103024.95+551622.7, \citealt{Rakshit:2018}) in addition to the $\gamma$-NLS1 FBQS~J1644+2619 \citep{Doi:2012}.  
However, according to the FR-I/II division in the diagram of 1.4-GHz luminosity versus the optical R-band magnitude \citep{Owen:1994,Ghisellini:2001}, \target~($L_{\rm 1.4GHz}=10^{24.7}$~W~Hz$^{-1}$ for the component N, $M_{R}=-25.1$~mag \citep{Mickaelian:2006}) is located in the FR-I region.  If we use the criterion of \citet{Landt:2006}, \target~is in the border region of FR~I/IIs ($L_{\rm 1.4GHz}=10^{24.5}$--$10^{26.0}$~W~Hz$^{-1}$).  
The jet power estimated by SED modelling for \target~is also at the lower end of the FSRQs \citep{Abdo:2009a}, which are thought to be FR~II radio galaxies viewed at small angles.    
\target~and these NLS1s may make edge-brightened radio structures as well as FR~IIs/FSRQs driven by high-mass-accretion-rate nuclei.  
In addition, the (presumably) much smaller core radii in the host galaxies of NLS1s may also allow jets to escape to kpc regions in the form of supersonic lobes even with black hole masses significantly lower than those of FR~IIs/FSRQs \citep{Doi:2012}.  However, FR-I morphology has rarely been found in NLS1.  This result may potentially be affected by a sensitivity-limited bias due to the low brightness of their extended structures.   
An FR-I morphology has been found in a nearby radio-quiet NLS1 with $L_{\rm 1.4GHz}\sim10^{23}$~W~Hz$^{-1}$ \citep[Mrk~1239;][]{Doi:2015}.  Such examples of NLS1s exhibiting FR~I/II radio structures suggest that the jet power is a key parameter of the kpc-scale jet activity.

The largest known projected sizes of large-scale radio structures in NLS1s are approximately 80 and 60~kpc on one side for SDSS~J110006.07+442144.3 \citep{Gabanyi:2018} and SDSS~J103024.95+551622.7 \citep{Rakshit:2018}, respectively.  Therefore, the separation of 60~kpc for component N in \target~is not an extreme case.  If a viewing angle of $3\degr$ is assumed \citep{Foschini:2011a,Foschini:2012}, a projected arm length of 60~kpc corresponds to a deprojected arm length of $\sim1$~Mpc, suggesting a huge radio galaxy.  The component N is almost perfectly aligned with the direction of the pc-scale jet in the sky plane (${\rm PA}\sim18\degr$).  This suggests that the direction of the jet axis has been stable for a long time, possibly at the viewing angle as well.    
These examples indicate the capability of some NLS1 central engines to maintain sufficient jet powers and lifetimes to develop huge radio galaxies up to Mpc scales.  
\target~might be a good example of an extreme NLS1 on an evolutionary path to an FSRQ as a more evolved source.

\subsection{The kinematic age?}\label{section:discussion:age}
The advancing speeds of the radio lobes at the hundred kpc and Mpc scales are significantly decelerated, generally down to $\beta \sim 0.01$, where $\beta$ is the advancing speed in units of speed of light \citep[e.g.][and references therein]{Kawakatu:2008}.   
The arm length of $\sim 1$~Mpc implies the jet activity for $\sim 3 \times 10^8$~year.  Such long-lasting activity contradicts the general argument for the lifetime of NLS1s.  A lifetime of $\sim10^7$~year is postulated for the NLS1 activity from the perspective of the number fraction of NLS1s ($\sim10$\%) in the AGNs ($\sim10^8$~year) and growth to supermassive black holes at the super-Eddington accretion rate \citep{Collin:2004,Kawaguchi:2004}.  

We discuss the possibility of a higher advancing speed and shorter kinematic age.  The observed asymmetry in two-sided radio structures can be generally understood as the result of the Doppler beaming effect, the light-travel-time effect, and/or hydrodynamic interactions with the inhomogeneous interstellar medium.  The first and second effects are useful for deriving the advancing speed of jets/lobes, although an intrinsic symmetry of the luminosity and a constant advancing speed on each side are necessary.  The third effect potentially accumulates, and then, the intrinsic symmetry would be severely broken at very large scales, such as $>100$~kpc.  
On the contrary, most FR~II-like radio structures in NLS1s simultaneously show moderate asymmetry in the flux density and arm length (e.g. FBQS~J1644+2619, SDSS~J120014.08$-$004638.7, \citealt{Doi:2012}; J0953+2836, J1435+3131 and J1722+5654, \citealt{Richards:2015} and SDSS~J103024.95+551622.7, \citealt{Rakshit:2018}), which is favourable for the Doppler beaming effect and the light-travel-time effect.  That is, the approaching jet side simultaneously exhibits a mildly higher flux density and a mildly larger arm length compared to those of the counter jet side.  Such observed properties suggest that the intrinsic symmetry of jets is reasonably maintained with $\beta \sim 0.1$--$0.3$ (at viewing angles of $<45\degr$), that is, for moderate advancing speeds at kpc scales.

In the following discussion, we estimate the advancing speed and kinematic age of the large-scale radio component in \target~on the basis of the Doppler beaming effect and the light-travel-time effect \citep[e.g.][]{Ghisellini:1993}, assuming that intrinsic symmetry with the luminosity and a constant speed are maintained at large scales.  The constraint for the flux ratio of the northern to southern components is $R_{\rm F} > 8.9$ (Section~\ref{section:results:kpcscaleradioemission}).  This leads to $\beta > 0.27$ from the Doppler beaming effect.  The apparent advancing speed on the sky would be $\beta_{\rm app} \ga 0.02$ if we assume a viewing angle of $3\degr$ \citep{Foschini:2011a,Foschini:2012}.  As a result, the kinematic age would need to be $t <1 \times 10^7$~year to develop the observed projected separation 60~kpc to the component N.  

This kinematic age is consistent with the lifetime $\sim10^7$~year that is generally considered to be the NLS1 phase.  Therefore, the radio component N possibly traces the recent jet producing activity of the NLS1 nucleus in PMN~J0948+0022.  
Similar advancing speeds and kinematic ages have also been estimated for NLS1s showing FR~II-like radio structures ($\beta\sim0.27$ and $t\ga 4\times 10^6$~year for $\gamma$-NLS1 FBQS~J1644+2619, \citealt{Doi:2012}; $\beta\sim0.03$--$0.3$ and $t \sim 10^6$--$10^7$~year for the three NLS1s in \citealt{Richards:2015} and $\beta\sim0.2$ and $t> 4\times 10^6$~year for the NLS1 SDSS~J103024.95+551622.7, \citealt{Rakshit:2018}).


\subsection{Physical conditions of the jet at $\sim 100$~pc scales}\label{section:discussion:physicalcondition}

The range of our investigation on the VLBA MOJAVE image of \target~is between 0.82 and 3.5~mas, corresponding to the deprojected distances between approximately 100 and 440~pc from the central engine, if we assume a viewing angle of $\theta = 3\degr$ \citep{Foschini:2011a,Foschini:2012}.  
The jet width profile in this region can be reproduced as $w_r \propto r^{1.12}$ (Section~\ref{section:results:pcscalejetstrcture}), indicating a nearly conical streamline for the jet.  This value is typical for AGN jets up to a few kpc; the $k$ values measured for 122~AGNs in the MOJAVE sample are distributed between approximately 0.5 and 1.3 with a median of 0.95 \citep{Pushkarev:2017}.  Mostly, the jet geometry is found to be close to that of conical ($k\sim1$) or quasi-parabolic ($0.5\la k <1$) streamlines.  

In addition, we obtain the profile of the flux density per unit length along the jet: $F_r^\prime \propto r^{-1.44}$.  Here, we examine the evolution of the internal energy in the jet plasma by using the obtained profiles of $w_r$ and $F_r^\prime$, in a manner similar to that used by \citet{Nakahara:2018}.  We assume pure (electron/positron) plasma in the jet.  
  
We define the ratio of the particle energy density to the magnetic energy density as $\xi_r \equiv u_{\rm e}/u_{\rm B}$, 
where $u_{\rm e}\ = \int^{\gamma_{\rm max}}_{\gamma_{\rm min}} K_0(r) \gamma^{1-p} m_{\rm e} c^2 d\gamma$ ($K_0(r)$, $p$ and $m_{\rm e}$ are the scale factors of electron density, the power-law index of the electron energy distribution and the rest mass of an electron, respectively) and $u_{\rm B}= B_r^2/8\pi$ ($B_r$ is the magnetic field strength).  
Synchrotron emissivity has a dependence of $\epsilon_\nu(r) \propto K_0(r) B_r^{(p+1)/2}$ \citep{Rybicki:1979}.  
The observed emission is Doppler-boosted, $F_r^\prime \propto \delta_r^{2-\alpha} \epsilon_\nu(r) w_r^2$, where $\delta_r$ and $\alpha$ are the Doppler factor and spectral index (Section~\ref{section:results:kpcscaleradioemission}), respectively.  
Finally, using the above relations, we derive the internal energy per unit length along the jet: $U_{\rm e}^\prime(r) \propto u_{\rm e}(r) w_r^2 = ( \xi_r^{p+1} w_r^{2(p+1)} F_r^{\prime 4} \delta_r^{-2(p+3)} )^{1/(p+5)}$.  
In the case of $\alpha=-0.7$ ($p=2.4$), by adopting $w_r \propto r^{1.12}$ and $F_r^{\prime} \propto r^{-1.44}$, we obtain $U_{\rm e}^\prime(r) \propto \xi_r^{0.46} \delta_r^{-1.46} r^{0.25}$.  This positive power-law index with $r$ indicates that the internal energy in the jet increases with distance.  The power-law index $d$, where $\delta_r \propto r^d$, is unknown\footnote{A weak dependence is presumably postulated; $d\sim-0.37$ is calculated based on the fitted deceleration from $\beta_{\rm app}\sim 19.1$ to $11.2$ at distances from 0.4 to 1.7~mas \citep{Lister:2016}.}    
but presumably $d < 0$ because a deceleration was detected for the superluminal component id2 for \target~in the MOJAVE monitoring \citep{Lister:2016}.  This deceleration makes $U_{\rm e}^\prime(r)$ increase more rapidly.  The power-law index $i$, where $\xi_r \propto r^i$, is also unknown but is presumably $i\ga0$ because the jet is initially magnetically dominated and eventually transforms to close to the equipartition condition or particle dominance as a global trend.  This behaviour also contributes to the increase in $U_{\rm e}^\prime(r)$ with distance.    
Therefore, the observed $r$-dependence of flux density in the conical streamline suggests that the investigated region ($\sim 100$--$440$~pc) is the site where the jet kinetic energy is converted into internal energy via some process of particle acceleration.


\subsection{Jet structure and supporting pressure}\label{section:discussion:pressure}

The inverse dependence $\Gamma \propto \phi_{\rm int}^{-1}$, where $\Gamma$ is the bulk Lorentz factor defined as $\Gamma \equiv (1-\beta^2)^{-0.5}$ and $\phi_{\rm int}$ is the intrinsic (full-)opening angle of the jet, is predicted via hydrodynamical \citep{Blandford:1979} and magnetic acceleration models \citep{Komissarov:2007} of relativistic jets.  
The parameter $\Gamma \phi_{\rm int}$ is related to the speed of the sideways expansion and the causal connection between the jet edge and its symmetry axis; $\Gamma \phi_{\rm int}/2 \la 1$ implies that the jet is causally connected \citep[e.g.][]{Clausen-Brown:2013}, and therefore, is sensitive to the boundary conditions between the jet and the surrounding medium.   

For \target, the apparent full-opening angle of the jet is $\phi_{\rm obs} = 41\degr \pm 1\degr$ on average (Section~\ref{section:results:pcscalejetstrcture}).  The deprojected full-opening angle is calculated to be $\phi_{\rm int} = 2.2\degr$ assuming $\theta = 3\degr$.  On the other hand, the two superluminal motions of $\beta_{\rm app} = 11.5 \pm 1.5$ for the component id2 and $\beta_{\rm app} = 5.5 \pm 1.9$ for id3 have been discovered at distances of $r\sim(0.4$--$1.7)$~mas and $r\sim(0.3$--$0.6)$~mas, respectively \citep{Lister:2016}.  Thus, the jet of \target~is still highly relativistic at the 100~pc scale.  Assuming $\theta = 3\degr$, $\Gamma=12.6$ and $7.9$ are calculated for these superluminal components (cf.~$\Gamma=10$ and $16$ at $r\sim0.03$~pc were adopted in the SED studies by \citealt{Abdo:2009a} and \citealt{Foschini:2011a}, respectively).  As a result, we obtain $\Gamma \phi_{\rm int}/2 = 0.24$ and $0.15$ for the jet of \target.  These values are reasonably consistent with the values estimated from large samples of AGNs: $\Gamma \phi_{\rm int}/2  = 0.17$ \citep{Jorstad:2005}, 0.13 \citep{Pushkarev:2009} and $\sim 0.2$ \citep{Clausen-Brown:2013} or $\Gamma \phi_{\rm int} = 0.35$ \citep{Pushkarev:2017}.  
Consequently, the jet of \target~is still causally connected, and then possibly influenced by the surrounding medium, at least up to the investigated distances between approximately 100 and 440~pc.

Next, we evaluate the internal pressure of the jet, which should be balanced with the external pressure of the surrounding medium.  Here, we discuss the expressions inherited from the formulas in the previous subsection (Section~\ref{section:discussion:physicalcondition}).  Assuming that the parameter $\xi_r$ is constant, the jet pressure can be represented as $p_{\rm jet} \propto u_{\rm e}$.  Similarly, we obtain $p_{\rm jet} \propto \delta_r^{-1.46} r^{-1.99}$.   The dependence of $\delta_r \propto r^d$ is unknown but presumably $d < 0$ because a deceleration was detected \citep{Lister:2016}.  Consequently, the jet outer boundary is thought to be supported by the external pressure with a dependence of $p_{\rm ext} \propto r^{-2}$ or shallower.    
At the external pressure of $p_{\rm ext} \propto r^{-2}$, the jet can be confined in width and collimated into a parabolic shape ($k=0.5$) as an acceleration zone \citep{Zakamska:2008,Tchekhovskoy:2008} in the inner region.  At the outer distances, the jet streamline reaches conical asymptotes ($k=1$) in the case of magnetically accelerated jets in magnetohydrodynamic models \citep{Komissarov:2009,Porth:2015}.  
The observed conical structure of \target~jet at the VLBI scales ($\sim100$--440~pc) is possibly equivalent to the predicted conical asymptotes at the relatively outer distance.  
Interestingly, 
in the nearest $\gamma$-NLS1 1H~0323+342 ($z=0.0629$), a jet width profile with a transition from quasi-parabolic ($k=0.58$) to outer conical/hyperbolic ($k=1.41$) streamlines \citep{Hada:2018} via a recollimation shock at a distance of $\sim160$~pc \citep{Doi:2018} has been revealed.  Thus, the distance range of the conical/hyperbolic streamline in 1H~0323+342 is similar to that in \target.  An acceleration zone in a parabolic streamline can be unresolved if it exists at small distances ($\la0.8$~mas or $\la100$~pc for \target), even though the conical jet structure at larger distances has been resolved.  The relativistic jets showing superluminal motions that have been observed in VLBI monitoring must have previously passed through this acceleration region.  

The presence of structural transitions appearing around the Bondi radius or the sphere of gravitational influence~(SGI) of a supermassive black hole has been discussed (\citealt{Stawarz:2002,Asada:2012,Tseng:2016}, cf.~\citealt{Hada:2018,Nakahara:2018,Nakahara:2019}).  The dependence of the pressure gradient may change at this radius.  Here, for \target, we estimated the radius of SGI, $r_{\rm SGI} = G M_{\rm BH} / \sigma_{\rm b}^2$ by using an empirical relation $M_{\rm BH} \approx 0.309 \times (\sigma_*/200~{\rm km~s^{-1}})^{4.38} 10^9 M_{\sun}$ \citep{Kormendy:2013}, where $G$, $M_{\rm BH}$, $\sigma_{\rm b}$ and $\sigma_*$ indicate the gravitational constant, mass of the black hole, velocity dispersion of the gravitationally bounded stars and the velocity dispersion of the stars in the central region, respectively.  We obtain the relation of $r_{\rm SGI} \propto M^{0.54}$ by assuming $\sigma_{\rm b} \approx \sigma_*$.  Assuming $M_{\rm BH} = 10^{8.2} M_{\sun}$ for \target~\citep{Abdo:2009}\footnote{The black hole mass for PMN~J0948+0022 has been estimated in a wide range from $M_{\rm BH} = 10^{7.5} M_{\sun}$ \citep{Yuan:2008} to $M_{\rm BH} = 10^{9.1} M_{\sun}$ \citep{Calderone:2013}, which leads to $r_{\rm SGI} \approx 10$--$70$~pc. }, 
we obtain an estimation of $r_{\rm SGI} \sim 23$~pc, equivalent to a projected angular scale of $\sim 0.2$~mas, which is much smaller than the angular resolution used in this study.  This can explain why we observed decelerating jets rather than the collimated acceleration region for this $\gamma$-NLS1 \target.  The derived pressure gradient in the observed jet implies that a host core radius with a flat pressure profile exists at $\sim20$--$100$~pc in this NLS1.     

Finally, we discuss the connection between the 100-pc scale jet and the large-scale radio emission N.  
The extension of the opening-angle profile ($\phi_{\rm obs} = 41\degr \pm 1\degr$ on average; Section~\ref{section:results:pcscalejetstrcture}) does not simply connect to the resolved size of emission N ($\la1\arcsec$ corresponding to an opening angle of $\la10\degr$; Section~\ref{section:results:kpcscaleradioemission}).  
A transition in the jet width profile is required at a distance between the VLBI and VLA scales.\footnote{An apparent recollimation at a large scale has been observed in Cygnus~A jets \citep{Nakahara:2019}.}  
A decrease in the internal pressure of the expanded jet can potentially result in a recollimation halfway to the component N.  Otherwise, the other inner layer of the spine/sheath jet structure \citep[e.g.][]{Ghisellini:2005} might reach to form the terminal shock as component N.

\section{Summary}\label{section:summary}
We investigated the jet structure of the archetype of $\gamma$-NLS1 PMN~J0948+0022 via direct radio imaging using the VLBA and VLA.  This study provides much more comprehensive jet properties than the previous studies at $\ga100$~pc and kpc scales, which is outside the dissipation region of $\gamma$ rays.  Our findings and their implications are as follows.
\begin{enumerate}
\item The VLA images that we generated from the archival data revealed a large-scale radio-emitting component, which is separated by 9.1~arcsec, corresponding to 60~kpc in the projected distance, at a position angle of $PA\sim18\degr$, which is in nearly the same direction as the pc-scale jet.  This component was resolved to an elongation of $PA\sim70\degr$ and showed a steep spectrum with an index of $\alpha \sim -1$.  These radio properties suggest a terminal shock in a radio lobe energized via the jet from the PMN~J0948+0022 nucleus.  

\item No counter component in the south direction was detected in our VLA images.  The flux-density ratio of the approaching lobe to the counter lobe indicates an advancing speed of $\beta> 0.27$ and a kinematic age of $<1\times10^7$~year up to $\sim1$~Mpc (if a viewing angle of $3\degr$ is assumed).  This timescale is consistent with the lifetime of the NLS1 activity ($\sim10^7$~year).  

\item The other radio emission located $71\arcsec$ northeast of PMN~J0948+0022 has been revealed by our VLA image at a high angular resolution to have a double-source morphology.  We concluded that this radio emission is a background source.  

\item The VLBA image from MOJAVE revealed the jet structure at distances ranging from 0.82 to 3.5~mas, corresponding to deprojected distances from approximately 100 to 440~pc, from the core, if we assume $\theta=3\degr$.  The jet width profile depending on the distance was fitted to $w_r\propto r^{1.12}$, that is, it is nearly conical.  The profile of the flux density per unit length along the jet is $F_r^\prime \propto r^{-1.44}$.  These dependences suggest that the internal energy in the jet increases with distance.  The investigated region ($\sim 100$--$440$~pc) of \target~is presumably the site where the jet kinetic energy is converted into internal energy.   

\item An apparent jet opening angle of $\sim41\deg$ was measured.  Considering this in conjunction with the previously reported superluminal motions, the jet flow of \target~is causally connected in the investigated distance range ($\sim 100$--$440$~pc).  Therefore, the nearly conical jet is likely supported by the pressure of the surrounding medium in the host galaxy of \target.    
\end{enumerate}

\section*{Acknowledgments} 
The Karl G.~Jansky Very Large Array and the Very Long Baseline Array are operated by the National Radio Astronomy Observatory, which is a facility of the National Science Foundation operated under cooperative agreement by Associated Universities, Inc.  
Funding for SDSS-III has been provided by the Alfred P. Sloan Foundation, the Participating Institutions, the National Science Foundation and the U.S. Department of Energy Office of Science. The SDSS-III web site is http://www.sdss3.org/.
SDSS-III is managed by the Astrophysical Research Consortium for the Participating Institutions of the SDSS-III Collaboration, including the University of Arizona, the Brazilian Participation Group, Brookhaven National Laboratory, Carnegie Mellon University, University of Florida, the French Participation Group, the German Participation Group, Harvard University, the Instituto de Astrofisica de Canarias, the Michigan State/Notre Dame/JINA Participation Group, Johns Hopkins University, Lawrence Berkeley National Laboratory, Max Planck Institute for Astrophysics, Max Planck Institute for Extraterrestrial Physics, New Mexico State University, New York University, Ohio State University, Pennsylvania State University, University of Portsmouth, Princeton University, the Spanish Participation Group, University of Tokyo, University of Utah, Vanderbilt University, University of Virginia, University of Washington and Yale University.  
This research used data from the MOJAVE database, which is maintained by the MOJAVE team (\citealt{Lister:2018a}, ApJS, 232, 12).




\label{lastpage}
\end{document}